%% file: main.tex
\documentclass[conference]{IEEEtran}

\usepackage[
	style=numeric,
	datamodel=software, 
	abbreviate=false,
	natbib=true,
	sorting=nyt,
	backend=biber,
	bibencoding=utf8,
	giveninits=true,
	url=true,
	doi=true,
	maxcitenames=10,
	maxbibnames=100]{biblatex}

\usepackage{software-biblatex}

\bibliography{biblio}

\usepackage{subcaption}
\usepackage{amsmath}
\usepackage{booktabs}
\usepackage{algpseudocode}
\usepackage{algorithm}
\usepackage{tikz}
\usepackage{microtype}
\usepackage{flushend}

\newcommand\blfootnote[1]{%
  \begingroup
  \renewcommand\thefootnote{}\footnote{#1}%
  \addtocounter{footnote}{-1}%
  \endgroup
}

\makeatletter
\let\blx@rerun@biber\relax
\makeatother

\begin{document}

\title{Sniffer deployment in urban area for human trajectory reconstruction and
contact tracing}
\author{\IEEEauthorblockN{Antoine Huchet}
\IEEEauthorblockA{\textit{L3i, La Rochelle University} \\
\textit{La Rochelle, France}\\
antoine.huchet@univ-lr.fr}
\and
\IEEEauthorblockN{Jean-Loup Guillaume}
\IEEEauthorblockA{\textit{L3i, La Rochelle University} \\
\textit{La Rochelle, France}\\
jean-loup.guillaume@univ-lr.fr}
\and
\IEEEauthorblockN{Yacine Ghamri-Doudane}
\IEEEauthorblockA{\textit{L3i, La Rochelle University} \\
\textit{La Rochelle, France}\\
yacine.ghamri@univ-lr.fr}
}

\maketitle

\begin{abstract}

To study the propagation of information from individual to individual, we need
mobility datasets. Existing datasets are not satisfactory because they are too
small, inaccurate or target a homogeneous subset of population. To draw valid
conclusions, we need sufficiently large and heterogeneous datasets. Thus we aim
for a passive non-intrusive data collection method, based on sniffers that are
to be deployed at some well-chosen street intersections. To this end, we need
optimization techniques for efficient placement of sniffers. We introduce a
heuristic, based on graph theory notions like the vertex cover problem along
with graph centrality measures.

\blfootnote{This work has been partially funded by the ANR MITIK project,
French National Research Agency (ANR), PRC AAPG2019.}
\end{abstract}


\section{Introduction}

Understanding the diffusion of information from individual to individual can
provide insights into various fields ranging from the study of opportunistic
networks to the spread of diseases. The spread of information or diseases occurs 
when individuals find themselves in close proximity.  We therefore need to trace 
such contacts which can be infered from mobility data. Indeed, if one knows the 
trajectories of the individuals, one can infer then contacts between them, that 
is, a potential propagation.

The lack of usable mobility dataset makes the study of propagation difficult.
Several active datasets are available like~\cite{calderoni2012location} but the
precision provided by the GPS is not sufficient to infer accurate contacts.
Sometimes due to measurements errors, an individual may be in a different street
than what is reported. Other datasets like~\cite{akkaya2015iot} only provide
small scale data because the data collection is restricted to indoor
environments. Other are available but they use an intrusive approach. This
implies that only the people that agreed to to be traced can be monitored, 
thus creating a bias in the data because the recruted population is often 
homogeneous. This usually corresponds to students at a computer science
department or participants at a conference as in~\cite{hui2005pocket}. None of
these solutions scale easily so they yield only small scale data.

\begin{figure}
\begin{center}
\includegraphics[width=0.653\linewidth]{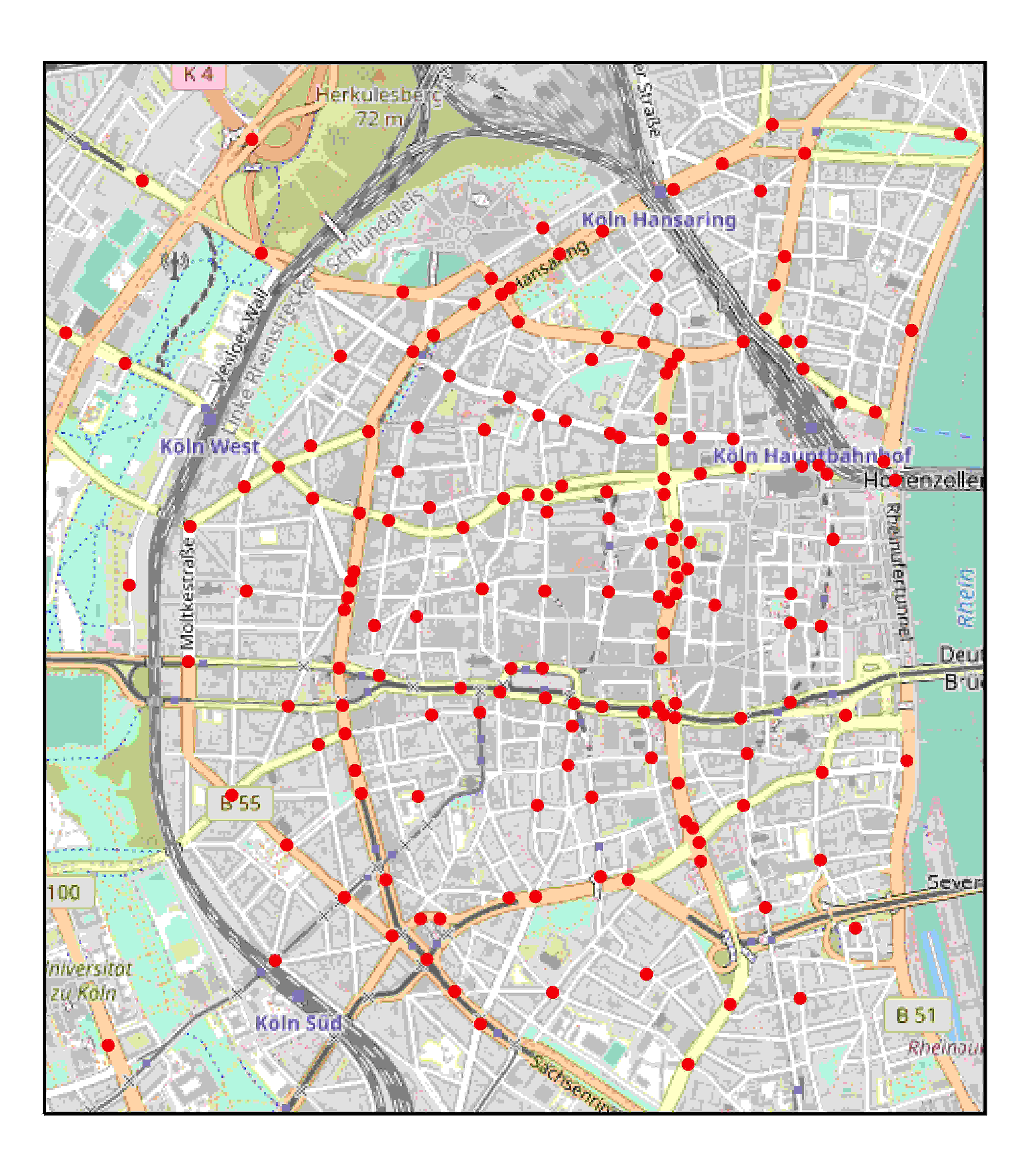}
\end{center}
\caption{map of downtown Cologne with sniffer deployment example}\label{map1}
\end{figure}

In this paper, we choose to study a passive non-intrusive data collection
approach because it can scale easily, provides sufficiently precise data, and
may target a heterogeneous population as individuals don't need to actively
report their location. We choose to work with sniffers that detect an
individual passing within a certain range. We can not get a full coverage of
our target area with our set of sniffers as the cost would be prohibitive, or
the target area too small to be relevant. We therefore need to develop
optimization techniques to efficiently place such sniffers. To do so, we
propose in this paper, a technique based on graph theory notions like the
vertex cover problem, along with graph centrality measures. The vertex cover
problems allows us to cover each edge, that is, each street section. Then the
graph centralities help us pick the most relevant street intersections. To
evaluate our solution, we retrieve the road graph of a city for which user
trajectories are known, then compute sniffer locations. Then we can measure the
performance of our solution by computing how many trajectories are covered. Our
experiments show that, on a realistic dataset, monitoring $20\%$ of the street
intersections of our target area, we are able to ``see'' about $95\%$ of the
trajectories enough times (i.e.\ multiple times with regards to the
trajectory). In order to illustrate the targeted size, a possible deployment is
shown in figure~\ref{map1}, where the red dots represents street intersections
where sniffers are deployed.

The remainder of this article is organized as follows: first we describe our
system model in section~\ref{sec:model}, then we give a brief reminder on the 
tools in section~\ref{sec:background} that are used in this article, then we 
review the related works in section~\ref{sec:related}. In 
section~\ref{sec:placement}, we explain our heuristic, then in 
section~\ref{sec:experiments}, we list the results based on a realistic mobility 
dataset.

\section{System model}
\label{sec:model}

\subsection{Road Networks and trajectories}

\begin{figure}
\begin{subfigure}{.2445\textwidth}
\begin{center}
\includegraphics[width=0.95\linewidth]{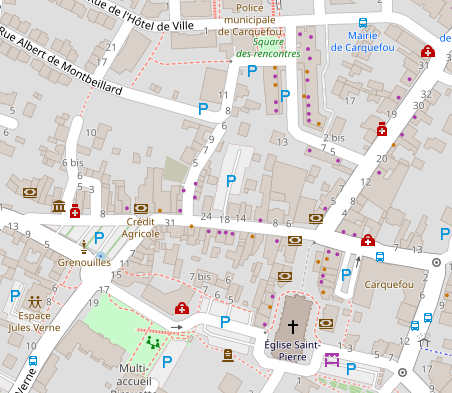}
\end{center}
\caption{Underlying map}\label{fig:map}
\end{subfigure}%
\begin{subfigure}{.2445\textwidth}
\begin{center}
\includegraphics[width=0.95\linewidth]{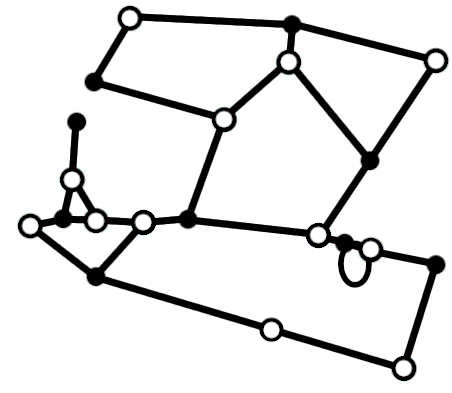}
\end{center}
\caption{A possible vertex cover}\label{fig:vertex_cover}
\end{subfigure}
\caption{Part of the Cologne road network (left), along with a road graph that models it (right).}\label{fig:maps}
\end{figure}

A graph $G$ consists of a set of vertices $V$ and a set of edges $E \subseteq V
\times V$. A graph can represent many things, including road networks: the edges 
would be street sections and the vertices would be the street intersections. 
Figure~\ref{fig:maps} shows an example of a road graph drawn next to its 
corresponding map.

Working with unweighted distances might not always capture human mobility
accurately as an edge can represent either a long or a short street section.
Since our road graph represents real world data, it's not difficult to enrich
it with additional information such as the GPS coordinates of our vertices.
This allows us to retrieve the real embedding of our graphs, thus to compute
the real world distances as the crow flies between vertices.

The map of the targeted area can be retrieved from OpenStreetMap and easily
converted into a graph as on figure~\ref{fig:map}. Note that this is a
simplified example as more vertices could be added as well as some self loops.
Such a representation is not unique. Working with a graph representation allows
us to take advantage of a wide range of algorithms and properties already known
about graphs.

A trajectory $T$ is a path in our road graph, i.e., an ordered list of
consecutive edges:

\[T = \{((u_0,u_1),(u_1,u_2), \ldots, (u_{k-1},u_k)), \forall i, (u_{i-1},u_i) \in E\}\]

The length of a trajectory in an unweighted graph is the number of edges
along the trajectory: here the length is $k$.

\subsection{Problem statement}

Our objective is to monitor the least amount of vertices of a road graph, while
maximizing the amount of trajectories seen by the sniffers, as well as the
amount of time they are seen. Our sniffers have a known range within which it 
is most likely that wireless activity will be captured. For this reason, we 
choose to work with the distance as the crow flies between the two endpoints of an 
edge.

We talk about vertices of a road graph rather than sniffers, because while we
aim at working with sniffers deployment in the end, the placement of sniffers
within a street intersection is another task in itself as it entails taking
into account the topology of the intersection. An intersection might require
several sniffers if it is too large or if there is too much traffic. This
constraints must be taken into account for a physical deployment of sniffers.
In this article, we restrict ourselves to the macroscopic view of sniffer
deployment by monitoring vertices in a graph.

\section{Background: Vertex Cover and graph centrality}
\label{sec:background}

Our objective is to select a subset of vertices that are evenly spread out 
in the graph. This can be done using a \emph{Vertex Cover} algorithm. Then, 
to reduce the amount of vertices, we select some of them based on some metrics 
called \emph{Centralities} measures. It has been shown that these measures 
allow an ``extended comprehension of the city structure, nicely capturing the 
skeleton of most central routes''~\cite{crucitti2006centrality}.

\subsection{Vertex Cover}

A \emph{Vertex Cover} is a subset of vertices of a graph that includes at 
least one endpoint of every edge of the graph. More formally, given a graph 
$G = (V,E)$, a \emph{Vertex Cover} is a $S \subseteq V$ such that $\forall 
(u,v) \in E$, $u \in S$ or $v \in S$. A \emph{Minimum Vertex Cover} is a set 
$S$ of minimum size.

Intuitively, if we put sniffers on every vertex of the vertex cover, then we
can monitor the whole graph, as every trajectory of length at least one has to
go through $S$. Sniffers need to be put where changes of direction happen,
because that's the information we need to reconstruct the shape of the
trajectory.

White nodes in Figure~\ref{fig:vertex_cover} are an example of a vertex 
cover: all edges of this graph have a white end. Note that this vertex cover
might not be minimum since there are edges for which ends are white. This 
problem belongs to the 21 problems that were proven NP-complete by Karp in 1972~\cite{karp1972reducibility}, therefore we often use approximation or 
heuristic algorithms.

In \emph{Maximum $k$-path Vertex Cover problem}~\cite{miyano2018complexity}, 
we are given $s$ sniffers, and we need to maximize the amount of paths of 
length at least $k$ that go through our at least one sniffer. 
Formally, given a graph $G = (V,E)$, an integer $k$ and an integer $s$, we 
want to find $S \subset V$, such that $\left|S\right| \leq s$, maximizing 
the amount of paths of length at least $k$ covered by $S$.

Few properties are known for this problem. It has been
shown~\cite{miyano2018complexity} that $MaxP_3VC \in$ \textsc{NP-Hard} for
split graphs, and \textsc{P} for graphs with bounded treewidth, $MaxP_3VC \in$
\textsc{XP} with respect to the treewidth ($tw$), $MaxP_3VC \in$ \textsc{FPT}
with respect to $s$ and $tw$. Finally, $MaxP_3VC \in$ \textsc{P} for trees.

\subsection{Graph centralities}
\label{subsec:centralities}

Centralities are values assigned to each node of a graph. They represent how
important a node is in a network. We review the main centrality measures from
the literature, which are all used in our solution.

\begin{enumerate}
    \item {\emph Degree centrality} is defined as the number of edges that 
    are adjacent to a given vertex~\cite{bavelas1948}.
    \item {\emph Strength centrality} of a vertex $u$ is the sum of the 
    weights of the adjacent edges of vertex $u$. This is the weighted 
    equivalent of the degree centrality and is also known as the weighted 
    degree centrality~\cite{barrat2004architecture}.
    \item {\emph Closeness centrality} of a vertex $u$ is the inverse of 
    the average length of the shortest path from vertex $u$ to all
    other vertices in the graph~\cite{bavelas1950communication}.
    \item {\emph Betweenness centrality} of a vertex $u$ is the number of 
    shortest paths which pass through $u$, divided by the number of 
    shortest paths between all pairs of vertices.~\cite{freeman1977set}
    \item {\emph Katz centrality} of a vertex $u$ takes into account the 
    amount of paths of length $k$ starting from $u$, while favouring short
    paths~\cite{katz1953new}.
    \item {\emph Eigenvector centrality} sums over the neighbours of a 
    vertex just like the degree centrality. It goes a step further by 
    weighting the neighbours based on their own eigenvector centrality. This
    way, being adjacent to central neighbours makes the vertex more central~\cite{newman2008mathematics}.
    \item {\emph Information centrality} is based on the ``information'' 
    contained in all possible paths of vertices~\cite{stephenson1989rethinking}.
    \item {\emph Accessibility} of a vertex $u$ is based on the set of 
    vertices that are reachable by self-avoiding walks of length $h$, 
    starting from $u$~\cite{viana2012effective}.
    \item {\emph Expected Force} aims at modeling the force of infection of a vertex~\cite{lawyer2015understanding}.
\end{enumerate}

\section{Related Work: Understanding human mobility in urban areas through
mobile technologies}
\label{sec:related}

There have been different approaches in the literature that tried to use
personal mobile devices in order to understand human mobility. Below, we
outline the main approaches and their limitation.

Mobility has been studied through an Android application
in~\cite{calderoni2012location}, or through people carying devices
in~\cite{hui2005pocket}. Those datasets are necessarily restricted by
the limited number of people involved in the study, and usually biased as they
only target a homogenous subset of people.

K. Akkaya \textit{et al.} used existing Wi-Fi networks inside buildings to
determine which rooms were occupied, based on Received Signal Strength
Indication (RSSI) and Media Access Control (MAC)
addresses~\cite{akkaya2015iot}. This is difficult to generalize to outdoor
settings as we can not leverage existing Wi-Fi networks in urban areas because
public Wi-Fi networks are not broadly available. Another downside is that this
only allows to monitor devices connected to a particular Wi-Fi network.

\cite{akhter2019iot} presents a pedestrian counting method based on Fresnel
lenses. These lenses leverage the infrared technology to count the
pedestrians. This study only counts users, it does not identify them so
trajectories cannot be reconstructed and contacts between users cannot be
inferred. Plus, the placement strategy is not discussed.

Sniffer placement has been studied in~\cite{ji2007optimal}
and~\cite{biaz2005impact}, at a small scale only. Those papers study the impact
of sniffer deployment on the localization accuracy, which is computed based on
the RSSI. Other studies like~\cite{chourasia2021wi}
and~\cite{grumert2018traffic} leverage the Wi-Fi technology to monitor the
speed and density of road traffic. Those studies are conducted at a large
scale, but the sniffer placement strategy isn't discussed.

In our work, we aim at monitoring human mobility in urban areas in order to
allow trajectory reconstruction and contact tracing. Our aim is to avoid any
social bias while achieving this. We thus propose to study what an efficient
sniffer placement approach, i.e.\ a passive mobility monitoring approach, would
bring to achieve our human mobility monitoring objective.

\section{Sniffer placement for efficient trajectory reconstruction and contact
tracing}
\label{sec:placement}

As mentioned earlier, the idea we are following for sniffer placement is to
start by computing a vertex cover which provides us with a subset of vertices
through which all trajectories must go. Then to use the centralities to select 
the most relevant vertices.

\subsection{Vertex Cover}

There is a 2-approximation algorithm, which according
to~\cite{papadimitriou1998combinatorial}, has been discovered independently by
F. Gavril and M. Yannakakis.

In the \emph{planar} case, that is, when a graph can be embedded in a plane,
without having two edges crossing, the Bar-Yehuda and Even's 
algorithm~\cite{bar1982approximating} is a linear time $\frac{5}{3}$-approximation 
for the \emph{Minimum Vertex Cover Problem}. Its bound is
slightly better than the 2-approximation algorithm from F. Gavril and M. Yannakakis~\cite{papadimitriou1998combinatorial} that also runs in linear
time~\cite{bar1982approximating}.

Note that in a road graph, there can be bridges and other
constructions that make it non-planar but a road graph can not be
arbitrarily complex because of real world constraints. Even if our
road graphs are not necessarily planar, our tests show that, on our
road graphs, this algorithm outperforms Gavril and Yannakakis'
2-approximation algorithm.

\subsection{Sniffer Placement Heuristic}

Our heuristic is built on top of the vertex cover computed with \textsc{Bar-Yehuda}: 
we perform sniffer placement decisions using Algorithm~\ref{heuristic}. More 
precisely, \textsc{Bar-Yehuda} returns a vertex cover $VC$ and the most central node 
$u$ of $VC$ is kept while all the vertices close to $u$ are removed from $VC$.

In the algorithm, \textsc{centrality}  returns a vector of centralities for all 
vertices in the graph. That is, each node is assigned a value going from $0$ to 
$1$. Different centralities can be used, such as the ones presented 
in subsection~\ref{subsec:centralities}.

This heuristic also takes as parameters an integer $k$ that influences the distance 
within which the vertices are pruned from the vertex cover set. Higher values 
mean more vertices will be pruned, resulting in less street intersections being 
monitored. A balance can be found between the ratio of uncaught trajectories and 
the proportion of intersections where sniffers are deployed.
The distance within which the vertices are pruned not only depends on $k$ but also 
on the centrality of $u$ (see line 7 and 8 of the algorithm). The most central 
vertices will require less pruning since they probably correspond to denser zones 
where many trajectories are likely to be found. Our experiments show that
monitoring more closely areas where central vertices are found allows us to see
more trajectories.

\begin{algorithm}
\caption{Sniffer Placement Heuristic (G, k)}\label{heuristic}
\begin{algorithmic}[1]
\State{$VC \gets \textsc{Bar-Yehuda(G)}$}
\State{$VC_k \gets \emptyset{}$}
\State{cent\_vector $\gets \textsc{centrality}(G)$}
\While{$VC \neq \emptyset$}
	\State{Pick the most central node $u$ from $VC$}
	\State{$VC_k \gets u$}
	\State{$c \gets 1-\left(\frac{\textrm{index of }u\textrm{ in
}\mathit{cent\_vector}}{\left|V\right|}\right)$}
	\State{Remove from $VC$, $u$ and all the vertices within distance
$\textsc{round}(k \times c)$ of $u$}
\EndWhile{}
\State{\Return{$VC_k$}}
\end{algorithmic}
\end{algorithm}

The implementation of these two algorithms is available on~\cite{heuristic}.

\section{Experiments on real datasets}\label{Results}
\label{sec:experiments}

\subsection{Datasets}

Two datasets are used to analyse the performance of our sniffer placement 
strategy.

\subsubsection{Cologne}

This dataset is provided by the TAPAS Cologne project, which
aims at reproducing, with the highest level of realism possible, 
car traffic in the greater urban area of the city of Cologne, Germany.
The resulting synthetic trace covers a region of 400 square kilometers for a
period of 24 hours in a typical working day. It includes 1 538 464
trajectories.~\cite{uppoor2013generation}
The dataset also includes a map of Cologne.

The map and the trajectories have been extracted with a tool that converts 
the map into a graph, and the trajectories into an ordered list of edges.
It is available on~\cite{sumo_extraction}.

Since the area included in the TAPAS Cologne project is wider than any
realistic deployment, and not homogeneous due to its different areas (urban,
non urban, residential) we restrict ourselves to downtown Cologne. We therefore 
shrink the data to a fourteen square kilometres area, corresponding to what is 
shown in figure~\ref{map1}

\begin{figure}
\begin{center}
\scalebox{0.9}{
\input{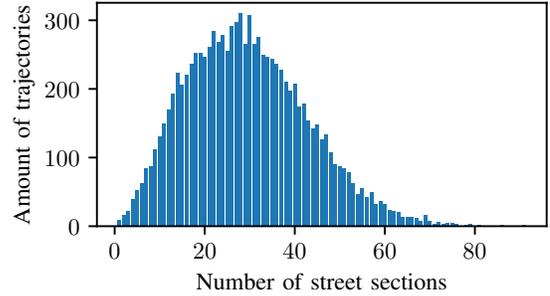}}
\end{center}
\caption{Distribution of the lengths, in number of street sections, of 
trajectories in downtown Cologne}\label{fig:lengthsdowntown}
\end{figure}

Our restricted dataset contains a road graph with 1080 vertices and 1615 edges,
along with 51 695 trajectories that never leave the restricted graph. The length
distribution is shown in figure~\ref{fig:lengthsdowntown}.

\subsubsection{Bologna}

The Bologna dataset was developed as part of the project iTRETRIS~\cite{iTETRIS}. 
This is a smaller dataset than Cologne, revolving around two main streets of the
city of Bologna. It simulates dense pedestrian movements as observed around 
football stadiums during big events such as football games or 
concerts~\cite{bieker2015traffic}. The road graph of this dataset contains 159
vertices, 215 edges and 11000 trajectories.

\subsubsection{Evaluation Methodology}

In the future, our goal is to reconstruct user trajectories from incomplete
measurements. To this end, we need to adapt our deployment of sniffers in order
to maximize the amount of trajectories that we can reconstruct with enough
precision. We consider that a trajectory can be reconstructed if it is observed 
"often" enough. In what follows, we consider that a trajectory is \emph{lost} 
and cannot be reconstructed, using two definitions: (i) when less than $x\%$ of 
its vertices are monitored or (ii) when it crosses less than $y$ monitored 
vertices. In the following we use the following values: $x=20\%$, $x=10\%$, 
$y=4$. These thresholds are purely arbitrary and their impact shall be 
studied in the future.

\subsubsection{Efficiency}\label{Efficiencysssec}

Let us now define a measure of efficiency. The goal is to minimize the number
of street intersections that we monitor, that is, use the least amount of
sniffers, along with minimizing the number of \emph{lost} trajectories. See
equation~\ref{efficiency}, where $S$ is the number of sniffers, $V$ is the 
number of vertices in our graph, $L$ is the number of \emph{lost} 
trajectories, and $T$ is the number of trajectories in our dataset. We need 
to maximize this function. This function ranges from $0$ to $1$, with $0$ 
being the worst and $1$ being the best.

\begin{equation}
\textrm{Efficiency} = \frac{|V| - |S|}{|V|} \times \frac{|T| - |L|}{|T|}
\label{efficiency}
\end{equation}

\begin{figure}
\begin{center}
\scalebox{0.9}{
\input{figures/choosingk.pgf}}
\end{center}
\caption{Efficiency as a function of $k$, for downtown Cologne, with expected
force centrality.}\label{fig:choosingk}
\end{figure}
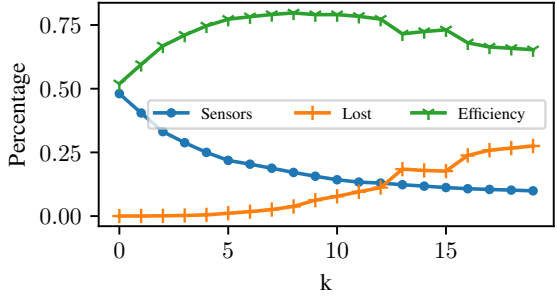

The $k$ value influences the distance within which the vertices
are pruned from the vertex cover set. As $k$ grows, the number of sniffers 
decreases, thus the number of \emph{lost} trajectories increases. $k$ should
be chosen in order to maximize the efficiency, as shown in
Figure~\ref{fig:choosingk}.

Note that table~\ref{table1} shows the $k$ value that maximizes the efficiency,
which is easy to do since we have a dataset with the trajectories. A way to
choose $k$ \emph{a priori} should be developed.

\subsubsection{Baselines}

To put our results into perspective, given that there is no similar work in
the state of the art, we also defined several more baselines to evaluate the
quality of our results.

The first baseline consists in selecting vertices at random. We choose the 
number of vertices that gives us the best results on average over 10 runs. 
Because we pick vertices at random, they should be evenly spread out across 
our graph. This corresponds to the line ``Random'' in table~\ref{table1}. We 
expect this method to yield poor results.

We used the \emph{strength} centrality using \emph{a priori} knowledge of
the trajectories: the weight of each edge is the number of trajectories that 
go through that edge. The \emph{strength} centrality of a node $u$ is then 
the sum of the weight of each of the edges that are adjacent to $u$. This 
centrality requires \emph{a priori} knowledge of the trajectories and is 
expected to outperforms other centralities.

We also implemented four greedy strategies that also use \emph{a priori} 
knowledge of the trajectories. ``Greedy Lost'' chooses, at each step, the 
unmonitored vertex through which the largest amount of \emph{lost} 
trajectories go, until $h$ vertices are chosen. We pick the $h$ that 
maximizes our efficiency function. ``Greedy No Lost'' does the same but 
stops when there is no more \emph{lost} trajectories. ``Greedy Traj'' chooses,
at each step, the unmonitored vertex through which the largest amount of 
trajectories go, until $h$ vertices are chosen. Finally, ``Greedy No Traj'' 
is the same but stops when $0\%$ of the trajectories are lost. 

\subsection{Results}

\subsubsection{Downtown Cologne dataset}

We run our heuristic on the road graph of downtown Cologne, with different
centrality measures in order to compare their efficiency.

Our results are summarized in table~\ref{table1}. The $k$ value is chosen in
order to maximize the efficiency function, as explained in
Section~\ref{Efficiencysssec}.

\begin{table*}
\begin{center}
\begin{tabular}{l |rrrr |rrrr |rrrr}
\emph{Lost} & \multicolumn{4}{|c|}{20\%} & \multicolumn{4}{|c|}{10\%} & \multicolumn{4}{|c}{4} \\
Centrality & k & Sniffers & Lost & Efficiency & k & Sniffers & Lost & Efficiency & k & Sniffers & Lost & Efficiency\\
\midrule
Strength        &  9  & 16.14\% &  4.10\% & 0.804 & 17  & 10.07\% &  4.14\% & 0.862 & 9   & 16.14\% &  8.84\% & 0.764\\
Greedy Traj     & & 17.13\% &  7.77\% & 0.764 & & 13.89\% &  5.54\% & 0.814 & & 17.13\% & 10.03\% & 0.746\\ 
Greedy No Traj  & & 86.02\% &     0\% & 0.140 & & 82.50\% &     0\% & 0.172 & & 97.96\% & 0.60\% & 0.019\\ 
Greedy Lost     & & 16.76\% &  8.54\% & 0.761 & & 10.83\% &  8.84\% & 0.813 &  & 17.41\% & 10.06\% & 0.743\\ 
Greedy No Lost  & & 85.46\% &     0\% & 0.144 & & 82.13\% &     0\% & 0.177 & & 97.96\% &  0.60\% & 0.019\\ 
\midrule
Degree          &  \textbf{5} & \textbf{16.48\%} & \textbf{4.54\%} & \textbf{0.797} & 7 & 13.80\% &  1.31\%  & 0.851 & 4 & 20.56\% & 7.29\% & 0.736\\
Katz            &   8 & 18.15\% &  3.00\% & 0.794 &  \textbf{20} &  \textbf{9.54\%} & \textbf{4.39\%} & \textbf{0.865} &   8 & 18.15\% &  9.21\% & 0.743\\
Expected Force  &   9 & 17.13\% &  4.14\% & 0.794 &  \textbf{23} &  \textbf{8.98\%} & \textbf{5.00\%} & \textbf{0.865} &   9 & 17.13\% & 10.70\% & 0.740\\
Accessibility   &  10 & 15.93\% &  5.88\% & 0.791 &  14 & 12.31\% & 2.50\% & 0.856 &   \textbf{9} & \textbf{17.31\%} &  \textbf{9.82\%} & \textbf{0.756}\\
Betweenness     &   7 & 20.56\% &  3.88\% & 0.764 &  13 & 12.96\% & 3.26\% & 0.842 &   7 & 20.56\% &  9.02\% & 0.723\\
Information     &   6 & 20.37\% &  4.21\% & 0.763 &  14 & 11.02\% & 4.79\% & 0.847 &  10 & 14.63\% & 13.52\% & 0.738\\
Closeness       &   4 & 23.33\% &  7.10\% & 0.712 &   9 & 13.61\% & 8.35\% & 0.792 &   4 & 23.33\% & 10.96\% & 0.683\\
Eigenvector     &   3 & 27.41\% &  3.39\% & 0.701 &   5 & 19.72\% & 3.19\% & 0.777 &   4 & 22.96\% & 11.32\% & 0.683\\
\midrule
Random          & 286 & 26.48\% &  5.60\% & 0.696 & 204 & 18.89\% & 4.10\% & 0.780 & 258 & 23.89\% & 14.12\% & 0.654\\
\end{tabular}
\caption{Results for each centrality, on downtown Cologne, considering different \emph{lost} thresholds.}\label{table1}
\end{center}
\end{table*}
\begin{figure}
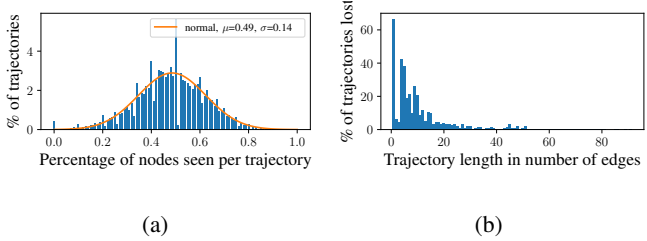

\begin{subfigure}{.2445\textwidth}
\begin{center}
\scalebox{0.41}{
\input{figures/cologne_downtown_new_def_20_weighted_deg_9_percent_lost_nc.pgf}}
\end{center}
\caption{}\label{fig:c20wd9c}
\end{subfigure}%
\begin{subfigure}{.2445\textwidth}
\begin{center}
\scalebox{0.41}{
\input{figures/cologne_downtown_new_def_20_weighted_deg_9_lost_by_length_nc.pgf}}
\end{center}
\caption{}\label{fig:c20wd9a}
\end{subfigure}
\caption{Cologne Downtown, $\emph{lost} = 20\%$, strength, $k=9$}\label{fig:c20wd9}
\end{figure}

The first column of table~\ref{table1} considers that a trajectory is lost when 
less than $20\%$ of the street intersections it crosses are monitored with 
sniffers. In this column, with most centralities, less than $20\%$ of the 
vertices need to be monitored to catch more than $95\%$ of the trajectories. 
This can also be seen on figure~\ref{fig:c20wd9c} that show how well observed 
are the trajectories in terms of monitored nodes.

A deeper inspection shows that the \emph{lost} trajectories are mostly short 
ones. The shorter a trajectory, the more unlikely it becomes to monitor $20\%$ 
of its vertices. For instance we have to put sniffers at every intersection to 
catch trajectories of length lower than $5$ and we can miss only one 
intersection for trajectories of length lower than $10$.
This is confirmed by Figure~\ref{fig:c20wd9a} that represents the amount of 
\emph{lost} trajectories as a function of their length. We can see that almost 
$40\%$ of the trajectories of length $4$ are \emph{lost} and more generally 
shorter trajectories tend to have more \emph{lost} among them.

\subsubsection{Impact of \emph{Lost}}

To better understand the impact of the definition of \emph{lost}, we also
consider a trajectory \emph{lost} when less than $10\%$ of the street
intersections it crosses are monitored. The second column of table~\ref{table1}
shows the results we get when taking into account this new definition. The
value of $k$ that maximizes our efficiency function is higher than on
table~\ref{table1}.  This is not surprising as we need less vertices to
consider a trajectory ``seen''. A higher value of $k$ means that there will be
less sniffers. Note that the percentage of ``lost'' remains in the same range.
With less sniffers and the same amount of \emph{lost} trajectories, the
efficiency function gives higher values.

\begin{figure}
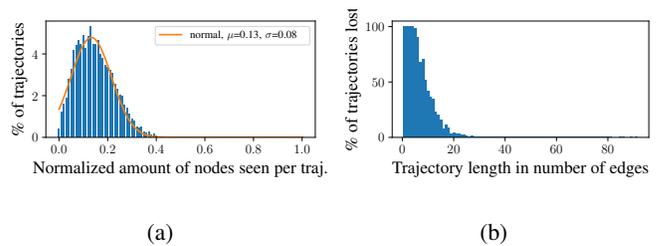

\begin{subfigure}{.2445\textwidth}
\begin{center}
\scalebox{0.41}{
\input{figures/cologne_downtown_old_def_4_weighted_deg_9_percent_lost_nc.pgf}}
\end{center}
\caption{}\label{fig:c4wd9c}
\end{subfigure}%
\begin{subfigure}{.2445\textwidth}
\begin{center}
\scalebox{0.41}{
\input{figures/cologne_downtown_old_def_4_weighted_deg_9_lost_by_length_nc.pgf}}
\end{center}
\caption{}\label{fig:c4wd9a}
\end{subfigure}%

\caption{Cologne downtown, $\emph{lost} = 4$, strength, $k=9$}\label{fig:c4wd9}
\end{figure}

Let us now consider a trajectory \emph{lost} when we have seen less than $4$ of
its street intersections. The third column of table~\ref{table1} shows how our
results differ according to this definition. As can be seen in
figure~\ref{fig:c4wd9a}, short trajectories are inevitably
missed. This is because we can not see 4 times a trajectory of length less than
4. Short trajectories are \emph{lost} most of the time for the same reason that
it's hard to see 4 times a trajectory of length close to 4. We see a sharp
decrease in percentage of \emph{lost} as trajectory length increases. The value
of $k$ that maximizes our efficiency function remains in the same range, with a
higher percentage of \emph{lost} trajectories, resulting in lower values for
our efficiency function.

Note that our ``Greedy No Lost'' method picked almost $98\%$ of the
vertices of the dataset because it only picks vertices through which \emph{lost}
trajectories go. Some vertices aren't crossed by a \emph{lost} trajectory or
even a trajectory at all.

\subsubsection{Bologna dataset}

Table~\ref{tableB} shows the results we got for the Bologna dataset, where
comparable results are obtained.

The first two columns correspond to trajectories being lost when less than
$20\%$, and $10\%$, of the street intersections it crosses are monitored. In
these columns, we can see that the efficiency is similar to that of Cologne,
but the optimal values of $k$ are smaller. This means that we prune less 
vertices, thus using proportionally more sniffers. This is probably because 
of the size of the dataset: the area is considerably smaller (about 10 times), 
but it has only 5 times less trajectories. Therefore, leaving more unmonitored 
vertices increases the amount of \emph{lost} trajectories.

In the third column we observe poorer results, which are due to the size of
the dataset. The average trajectory length in Cologne is 30
(see Fig~\ref{fig:lengthsdowntown}) while it is 11 in Bologna. Thus it is
harder to see $4$ of their vertices.




\begin{table*}
\begin{center}
\begin{tabular}{l |rrrr |rrrr |rrrr}
\emph{Lost} & \multicolumn{4}{|c|}{20\%} & \multicolumn{4}{|c|}{10\%} & \multicolumn{4}{|c}{4} \\
Centrality      & k  & Sniffers & Lost   & Efficiency & k  & Sniffers & Lost  & Efficiency & k  & Sniffers & Lost   & Efficiency\\
\midrule
Strength        &  6 & 21.90\% &  2.40\% & 0.762 & 19 & 12.41\% & 1.50\% & 0.863 & 10 & 16.79\% & 43.28\% & 0.472\\
Greedy Lost     &    & 18.87\% &  9.05\% & 0.738 &    & 11.32\% & 8.03\% & 0.816 &     & 23.90\% & 27.84\% & 0.549\\
Greedy No Lost  &    & 45.28\% &     0\% & 0.547 &    & 40.25\% &    0\% & 0.597 &     & 84.28\% & 7.44\% & 0.140\\
Greedy Traj     &    & 18.87\% &  9.05\% & 0.738 &    & 11.32\% & 8.03\% & 0.816 &     & 23.90\% & 27.84\% & 0.549\\
Greedy No Traj  &    & 45.28\% &     0\% & 0.516 &    & 40.25\% &    0\% & 0.579 &     & 84.28\% &  7.44\% & 0.134\\
\midrule
Degree          &  \textbf{4} & \textbf{26.75\%} & \textbf{0.56\%} & \textbf{0.728} & 12 &  8.92\% & 3.82\% & 0.876 & \textbf{3} & \textbf{29.30\%} & \textbf{34.31\%} & \textbf{0.464}\\
Katz            &  4 & 27.39\% &  0.48\% & 0.723 & \textbf{19} &  \textbf{9.55\%} &  \textbf{0.90\%} & \textbf{0.896} &  3 & 29.94\% & 42.54\% & 0.403\\
Expected Force  &  4 & 27.39\% &  0.56\% & 0.722 & 23 &  8.91\% & 6.04\% & 0.856 &  6 & 19.11\% & 57.53\% & 0.344\\
Betweenness     &  3 & 29.30\% &  0.75\% & 0.702 & 13 & 13.38\% & 2.22\% & 0.847 &  8 & 17.20\% & 60.68\% & 0.326\\
Accessibility   &  3 & 30.57\% &  4.61\% & 0.662 &  5 & 22.93\% & 1.07\% & 0.762 &  3 & 30.57\% & 47.32\% & 0.366\\
Eigenvector     &  3 & 28.66\% & 10.48\% & 0.639 &  6 & 18.47\% & 6.31\% & 0.764 &  1 & 40.13\% & 36.62\% & 0.379\\
Closeness       &  3 & 27.39\% & 12.87\% & 0.633 &  5 & 22.29\% & 5.11\% & 0.737 &  2 & 31.21\% & 48.15\% & 0.357\\
Information     &  3 & 28.03\% & 18.74\% & 0.585 &  6 & 20.38\% & 8.05\% & 0.732 &  2 & 35.67\% & 45.14\% & 0.353\\
\midrule
Random          & 50 & 31.45\% & 10.61\% & 0.613 & 42 & 26.42\% & 6.42\% & 0.691 & 71 & 44.65\% & 33.30\% & 0.369\\
\end{tabular}
\caption{Results for each centrality, on downtown Bologna, considering different \emph{lost} thresholds.}\label{tableB}
\end{center}
\end{table*}

\section{Conclusion and future work}
\label{sec:conclusion}

We developed a method for efficient placement of sniffers in an urban network,
for trajectory reconstruction and contact tracing. This method takes advantage
of the Vertex Cover problem and the notion of graph centralities to pick the
most important vertices. We showed
that on our datasets, the Degree, Katz and Expected Force centralities provide
the best results. Considering that we need to see a trajectory in at least
$20\%$ of the street intersections it crosses, on our Cologne downtown dataset,
we showed that monitoring as little as $16\%$ of the street intersection, only
$5\%$ of the trajectories of the dataset were lost. To put our results into
perspective, we also compared our method with other methods that take advantage
of the solution. These are used as forms of upper bounds. We also developed a
naive solution to act as a form of lower bound. Our solution is on average
rather close to our upper bounds.

\subsection{Future Work}

We could refine our objective: instead of maximizing the amount of time a
trajectory is seen, we could aim at seeing it just enough to reconstruct it
with sufficient accuracy.

We should take into account the topology of each street intersection,
as we might need to place several sniffers at some intersections in order
to get a complete coverage of the intersection, making them more expensive to
monitor. Our heuristic could therefore take into account the cost of deployment
at each intersection.

The next steps would be to deploy our sniffers, with which we could gather our
own dataset of trajectories. Then we could reconstruct the users trajectories.

Once a first set of trajectories is obtained, we could then use these
trajectories to weigh our road graph and redeploy with the knowledge given by
these trajectories. Redeployments should help us converge to a better solution.

\printbibliography{}

\end{document}

%% file: figures/choosingk.pgf
\begingroup%
\makeatletter%
\begin{pgfpicture}%
\pgfpathrectangle{\pgfpointorigin}{\pgfqpoint{3.400000in}{2.000000in}}%
\pgfusepath{use as bounding box, clip}%
\begin{pgfscope}%
\pgfsetbuttcap%
\pgfsetmiterjoin%
\definecolor{currentfill}{rgb}{1.000000,1.000000,1.000000}%
\pgfsetfillcolor{currentfill}%
\pgfsetlinewidth{0.000000pt}%
\definecolor{currentstroke}{rgb}{1.000000,1.000000,1.000000}%
\pgfsetstrokecolor{currentstroke}%
\pgfsetdash{}{0pt}%
\pgfpathmoveto{\pgfqpoint{0.000000in}{0.000000in}}%
\pgfpathlineto{\pgfqpoint{3.400000in}{0.000000in}}%
\pgfpathlineto{\pgfqpoint{3.400000in}{2.000000in}}%
\pgfpathlineto{\pgfqpoint{0.000000in}{2.000000in}}%
\pgfpathlineto{\pgfqpoint{0.000000in}{0.000000in}}%
\pgfpathclose%
\pgfusepath{fill}%
\end{pgfscope}%
\begin{pgfscope}%
\pgfsetbuttcap%
\pgfsetmiterjoin%
\definecolor{currentfill}{rgb}{1.000000,1.000000,1.000000}%
\pgfsetfillcolor{currentfill}%
\pgfsetlinewidth{0.000000pt}%
\definecolor{currentstroke}{rgb}{0.000000,0.000000,0.000000}%
\pgfsetstrokecolor{currentstroke}%
\pgfsetstrokeopacity{0.000000}%
\pgfsetdash{}{0pt}%
\pgfpathmoveto{\pgfqpoint{0.603704in}{0.549691in}}%
\pgfpathlineto{\pgfqpoint{3.250000in}{0.549691in}}%
\pgfpathlineto{\pgfqpoint{3.250000in}{1.850000in}}%
\pgfpathlineto{\pgfqpoint{0.603704in}{1.850000in}}%
\pgfpathlineto{\pgfqpoint{0.603704in}{0.549691in}}%
\pgfpathclose%
\pgfusepath{fill}%
\end{pgfscope}%
\begin{pgfscope}%
\pgfsetbuttcap%
\pgfsetroundjoin%
\definecolor{currentfill}{rgb}{0.000000,0.000000,0.000000}%
\pgfsetfillcolor{currentfill}%
\pgfsetlinewidth{0.803000pt}%
\definecolor{currentstroke}{rgb}{0.000000,0.000000,0.000000}%
\pgfsetstrokecolor{currentstroke}%
\pgfsetdash{}{0pt}%
\pgfsys@defobject{currentmarker}{\pgfqpoint{0.000000in}{-0.048611in}}{\pgfqpoint{0.000000in}{0.000000in}}{%
\pgfpathmoveto{\pgfqpoint{0.000000in}{0.000000in}}%
\pgfpathlineto{\pgfqpoint{0.000000in}{-0.048611in}}%
\pgfusepath{stroke,fill}%
}%
\begin{pgfscope}%
\pgfsys@transformshift{0.723990in}{0.549691in}%
\pgfsys@useobject{currentmarker}{}%
\end{pgfscope}%
\end{pgfscope}%
\begin{pgfscope}%
\definecolor{textcolor}{rgb}{0.000000,0.000000,0.000000}%
\pgfsetstrokecolor{textcolor}%
\pgfsetfillcolor{textcolor}%
\pgftext[x=0.723990in,y=0.452469in,,top]{\color{textcolor}\rmfamily\fontsize{10.000000}{12.000000}\selectfont \(\displaystyle {0}\)}%
\end{pgfscope}%
\begin{pgfscope}%
\pgfsetbuttcap%
\pgfsetroundjoin%
\definecolor{currentfill}{rgb}{0.000000,0.000000,0.000000}%
\pgfsetfillcolor{currentfill}%
\pgfsetlinewidth{0.803000pt}%
\definecolor{currentstroke}{rgb}{0.000000,0.000000,0.000000}%
\pgfsetstrokecolor{currentstroke}%
\pgfsetdash{}{0pt}%
\pgfsys@defobject{currentmarker}{\pgfqpoint{0.000000in}{-0.048611in}}{\pgfqpoint{0.000000in}{0.000000in}}{%
\pgfpathmoveto{\pgfqpoint{0.000000in}{0.000000in}}%
\pgfpathlineto{\pgfqpoint{0.000000in}{-0.048611in}}%
\pgfusepath{stroke,fill}%
}%
\begin{pgfscope}%
\pgfsys@transformshift{1.357075in}{0.549691in}%
\pgfsys@useobject{currentmarker}{}%
\end{pgfscope}%
\end{pgfscope}%
\begin{pgfscope}%
\definecolor{textcolor}{rgb}{0.000000,0.000000,0.000000}%
\pgfsetstrokecolor{textcolor}%
\pgfsetfillcolor{textcolor}%
\pgftext[x=1.357075in,y=0.452469in,,top]{\color{textcolor}\rmfamily\fontsize{10.000000}{12.000000}\selectfont \(\displaystyle {5}\)}%
\end{pgfscope}%
\begin{pgfscope}%
\pgfsetbuttcap%
\pgfsetroundjoin%
\definecolor{currentfill}{rgb}{0.000000,0.000000,0.000000}%
\pgfsetfillcolor{currentfill}%
\pgfsetlinewidth{0.803000pt}%
\definecolor{currentstroke}{rgb}{0.000000,0.000000,0.000000}%
\pgfsetstrokecolor{currentstroke}%
\pgfsetdash{}{0pt}%
\pgfsys@defobject{currentmarker}{\pgfqpoint{0.000000in}{-0.048611in}}{\pgfqpoint{0.000000in}{0.000000in}}{%
\pgfpathmoveto{\pgfqpoint{0.000000in}{0.000000in}}%
\pgfpathlineto{\pgfqpoint{0.000000in}{-0.048611in}}%
\pgfusepath{stroke,fill}%
}%
\begin{pgfscope}%
\pgfsys@transformshift{1.990161in}{0.549691in}%
\pgfsys@useobject{currentmarker}{}%
\end{pgfscope}%
\end{pgfscope}%
\begin{pgfscope}%
\definecolor{textcolor}{rgb}{0.000000,0.000000,0.000000}%
\pgfsetstrokecolor{textcolor}%
\pgfsetfillcolor{textcolor}%
\pgftext[x=1.990161in,y=0.452469in,,top]{\color{textcolor}\rmfamily\fontsize{10.000000}{12.000000}\selectfont \(\displaystyle {10}\)}%
\end{pgfscope}%
\begin{pgfscope}%
\pgfsetbuttcap%
\pgfsetroundjoin%
\definecolor{currentfill}{rgb}{0.000000,0.000000,0.000000}%
\pgfsetfillcolor{currentfill}%
\pgfsetlinewidth{0.803000pt}%
\definecolor{currentstroke}{rgb}{0.000000,0.000000,0.000000}%
\pgfsetstrokecolor{currentstroke}%
\pgfsetdash{}{0pt}%
\pgfsys@defobject{currentmarker}{\pgfqpoint{0.000000in}{-0.048611in}}{\pgfqpoint{0.000000in}{0.000000in}}{%
\pgfpathmoveto{\pgfqpoint{0.000000in}{0.000000in}}%
\pgfpathlineto{\pgfqpoint{0.000000in}{-0.048611in}}%
\pgfusepath{stroke,fill}%
}%
\begin{pgfscope}%
\pgfsys@transformshift{2.623246in}{0.549691in}%
\pgfsys@useobject{currentmarker}{}%
\end{pgfscope}%
\end{pgfscope}%
\begin{pgfscope}%
\definecolor{textcolor}{rgb}{0.000000,0.000000,0.000000}%
\pgfsetstrokecolor{textcolor}%
\pgfsetfillcolor{textcolor}%
\pgftext[x=2.623246in,y=0.452469in,,top]{\color{textcolor}\rmfamily\fontsize{10.000000}{12.000000}\selectfont \(\displaystyle {15}\)}%
\end{pgfscope}%
\begin{pgfscope}%
\definecolor{textcolor}{rgb}{0.000000,0.000000,0.000000}%
\pgfsetstrokecolor{textcolor}%
\pgfsetfillcolor{textcolor}%
\pgftext[x=1.926852in,y=0.273457in,,top]{\color{textcolor}\rmfamily\fontsize{10.000000}{12.000000}\selectfont k}%
\end{pgfscope}%
\begin{pgfscope}%
\pgfsetbuttcap%
\pgfsetroundjoin%
\definecolor{currentfill}{rgb}{0.000000,0.000000,0.000000}%
\pgfsetfillcolor{currentfill}%
\pgfsetlinewidth{0.803000pt}%
\definecolor{currentstroke}{rgb}{0.000000,0.000000,0.000000}%
\pgfsetstrokecolor{currentstroke}%
\pgfsetdash{}{0pt}%
\pgfsys@defobject{currentmarker}{\pgfqpoint{-0.048611in}{0.000000in}}{\pgfqpoint{-0.000000in}{0.000000in}}{%
\pgfpathmoveto{\pgfqpoint{-0.000000in}{0.000000in}}%
\pgfpathlineto{\pgfqpoint{-0.048611in}{0.000000in}}%
\pgfusepath{stroke,fill}%
}%
\begin{pgfscope}%
\pgfsys@transformshift{0.603704in}{0.608499in}%
\pgfsys@useobject{currentmarker}{}%
\end{pgfscope}%
\end{pgfscope}%
\begin{pgfscope}%
\definecolor{textcolor}{rgb}{0.000000,0.000000,0.000000}%
\pgfsetstrokecolor{textcolor}%
\pgfsetfillcolor{textcolor}%
\pgftext[x=0.259568in, y=0.560274in, left, base]{\color{textcolor}\rmfamily\fontsize{10.000000}{12.000000}\selectfont \(\displaystyle {0.00}\)}%
\end{pgfscope}%
\begin{pgfscope}%
\pgfsetbuttcap%
\pgfsetroundjoin%
\definecolor{currentfill}{rgb}{0.000000,0.000000,0.000000}%
\pgfsetfillcolor{currentfill}%
\pgfsetlinewidth{0.803000pt}%
\definecolor{currentstroke}{rgb}{0.000000,0.000000,0.000000}%
\pgfsetstrokecolor{currentstroke}%
\pgfsetdash{}{0pt}%
\pgfsys@defobject{currentmarker}{\pgfqpoint{-0.048611in}{0.000000in}}{\pgfqpoint{-0.000000in}{0.000000in}}{%
\pgfpathmoveto{\pgfqpoint{-0.000000in}{0.000000in}}%
\pgfpathlineto{\pgfqpoint{-0.048611in}{0.000000in}}%
\pgfusepath{stroke,fill}%
}%
\begin{pgfscope}%
\pgfsys@transformshift{0.603704in}{0.979213in}%
\pgfsys@useobject{currentmarker}{}%
\end{pgfscope}%
\end{pgfscope}%
\begin{pgfscope}%
\definecolor{textcolor}{rgb}{0.000000,0.000000,0.000000}%
\pgfsetstrokecolor{textcolor}%
\pgfsetfillcolor{textcolor}%
\pgftext[x=0.259568in, y=0.930988in, left, base]{\color{textcolor}\rmfamily\fontsize{10.000000}{12.000000}\selectfont \(\displaystyle {0.25}\)}%
\end{pgfscope}%
\begin{pgfscope}%
\pgfsetbuttcap%
\pgfsetroundjoin%
\definecolor{currentfill}{rgb}{0.000000,0.000000,0.000000}%
\pgfsetfillcolor{currentfill}%
\pgfsetlinewidth{0.803000pt}%
\definecolor{currentstroke}{rgb}{0.000000,0.000000,0.000000}%
\pgfsetstrokecolor{currentstroke}%
\pgfsetdash{}{0pt}%
\pgfsys@defobject{currentmarker}{\pgfqpoint{-0.048611in}{0.000000in}}{\pgfqpoint{-0.000000in}{0.000000in}}{%
\pgfpathmoveto{\pgfqpoint{-0.000000in}{0.000000in}}%
\pgfpathlineto{\pgfqpoint{-0.048611in}{0.000000in}}%
\pgfusepath{stroke,fill}%
}%
\begin{pgfscope}%
\pgfsys@transformshift{0.603704in}{1.349926in}%
\pgfsys@useobject{currentmarker}{}%
\end{pgfscope}%
\end{pgfscope}%
\begin{pgfscope}%
\definecolor{textcolor}{rgb}{0.000000,0.000000,0.000000}%
\pgfsetstrokecolor{textcolor}%
\pgfsetfillcolor{textcolor}%
\pgftext[x=0.259568in, y=1.301701in, left, base]{\color{textcolor}\rmfamily\fontsize{10.000000}{12.000000}\selectfont \(\displaystyle {0.50}\)}%
\end{pgfscope}%
\begin{pgfscope}%
\pgfsetbuttcap%
\pgfsetroundjoin%
\definecolor{currentfill}{rgb}{0.000000,0.000000,0.000000}%
\pgfsetfillcolor{currentfill}%
\pgfsetlinewidth{0.803000pt}%
\definecolor{currentstroke}{rgb}{0.000000,0.000000,0.000000}%
\pgfsetstrokecolor{currentstroke}%
\pgfsetdash{}{0pt}%
\pgfsys@defobject{currentmarker}{\pgfqpoint{-0.048611in}{0.000000in}}{\pgfqpoint{-0.000000in}{0.000000in}}{%
\pgfpathmoveto{\pgfqpoint{-0.000000in}{0.000000in}}%
\pgfpathlineto{\pgfqpoint{-0.048611in}{0.000000in}}%
\pgfusepath{stroke,fill}%
}%
\begin{pgfscope}%
\pgfsys@transformshift{0.603704in}{1.720639in}%
\pgfsys@useobject{currentmarker}{}%
\end{pgfscope}%
\end{pgfscope}%
\begin{pgfscope}%
\definecolor{textcolor}{rgb}{0.000000,0.000000,0.000000}%
\pgfsetstrokecolor{textcolor}%
\pgfsetfillcolor{textcolor}%
\pgftext[x=0.259568in, y=1.672414in, left, base]{\color{textcolor}\rmfamily\fontsize{10.000000}{12.000000}\selectfont \(\displaystyle {0.75}\)}%
\end{pgfscope}%
\begin{pgfscope}%
\definecolor{textcolor}{rgb}{0.000000,0.000000,0.000000}%
\pgfsetstrokecolor{textcolor}%
\pgfsetfillcolor{textcolor}%
\pgftext[x=0.204012in,y=1.199846in,,bottom,rotate=90.000000]{\color{textcolor}\rmfamily\fontsize{10.000000}{12.000000}\selectfont Percentage}%
\end{pgfscope}%
\begin{pgfscope}%
\pgfpathrectangle{\pgfqpoint{0.603704in}{0.549691in}}{\pgfqpoint{2.646296in}{1.300309in}}%
\pgfusepath{clip}%
\pgfsetrectcap%
\pgfsetroundjoin%
\pgfsetlinewidth{1.505625pt}%
\definecolor{currentstroke}{rgb}{0.121569,0.466667,0.705882}%
\pgfsetstrokecolor{currentstroke}%
\pgfsetdash{}{0pt}%
\pgfpathmoveto{\pgfqpoint{0.723990in}{1.321093in}}%
\pgfpathlineto{\pgfqpoint{0.850607in}{1.208506in}}%
\pgfpathlineto{\pgfqpoint{0.977224in}{1.100038in}}%
\pgfpathlineto{\pgfqpoint{1.103841in}{1.035506in}}%
\pgfpathlineto{\pgfqpoint{1.230458in}{0.979213in}}%
\pgfpathlineto{\pgfqpoint{1.357075in}{0.932530in}}%
\pgfpathlineto{\pgfqpoint{1.483692in}{0.910562in}}%
\pgfpathlineto{\pgfqpoint{1.610310in}{0.887221in}}%
\pgfpathlineto{\pgfqpoint{1.736927in}{0.862507in}}%
\pgfpathlineto{\pgfqpoint{1.863544in}{0.840539in}}%
\pgfpathlineto{\pgfqpoint{1.990161in}{0.819943in}}%
\pgfpathlineto{\pgfqpoint{2.116778in}{0.806213in}}%
\pgfpathlineto{\pgfqpoint{2.243395in}{0.800721in}}%
\pgfpathlineto{\pgfqpoint{2.370012in}{0.791110in}}%
\pgfpathlineto{\pgfqpoint{2.496629in}{0.782872in}}%
\pgfpathlineto{\pgfqpoint{2.623246in}{0.774634in}}%
\pgfpathlineto{\pgfqpoint{2.749863in}{0.767769in}}%
\pgfpathlineto{\pgfqpoint{2.876480in}{0.763650in}}%
\pgfpathlineto{\pgfqpoint{3.003097in}{0.759531in}}%
\pgfpathlineto{\pgfqpoint{3.129714in}{0.755412in}}%
\pgfusepath{stroke}%
\end{pgfscope}%
\begin{pgfscope}%
\pgfpathrectangle{\pgfqpoint{0.603704in}{0.549691in}}{\pgfqpoint{2.646296in}{1.300309in}}%
\pgfusepath{clip}%
\pgfsetbuttcap%
\pgfsetroundjoin%
\definecolor{currentfill}{rgb}{0.121569,0.466667,0.705882}%
\pgfsetfillcolor{currentfill}%
\pgfsetlinewidth{1.003750pt}%
\definecolor{currentstroke}{rgb}{0.121569,0.466667,0.705882}%
\pgfsetstrokecolor{currentstroke}%
\pgfsetdash{}{0pt}%
\pgfsys@defobject{currentmarker}{\pgfqpoint{-0.020833in}{-0.020833in}}{\pgfqpoint{0.020833in}{0.020833in}}{%
\pgfpathmoveto{\pgfqpoint{0.000000in}{-0.020833in}}%
\pgfpathcurveto{\pgfqpoint{0.005525in}{-0.020833in}}{\pgfqpoint{0.010825in}{-0.018638in}}{\pgfqpoint{0.014731in}{-0.014731in}}%
\pgfpathcurveto{\pgfqpoint{0.018638in}{-0.010825in}}{\pgfqpoint{0.020833in}{-0.005525in}}{\pgfqpoint{0.020833in}{0.000000in}}%
\pgfpathcurveto{\pgfqpoint{0.020833in}{0.005525in}}{\pgfqpoint{0.018638in}{0.010825in}}{\pgfqpoint{0.014731in}{0.014731in}}%
\pgfpathcurveto{\pgfqpoint{0.010825in}{0.018638in}}{\pgfqpoint{0.005525in}{0.020833in}}{\pgfqpoint{0.000000in}{0.020833in}}%
\pgfpathcurveto{\pgfqpoint{-0.005525in}{0.020833in}}{\pgfqpoint{-0.010825in}{0.018638in}}{\pgfqpoint{-0.014731in}{0.014731in}}%
\pgfpathcurveto{\pgfqpoint{-0.018638in}{0.010825in}}{\pgfqpoint{-0.020833in}{0.005525in}}{\pgfqpoint{-0.020833in}{0.000000in}}%
\pgfpathcurveto{\pgfqpoint{-0.020833in}{-0.005525in}}{\pgfqpoint{-0.018638in}{-0.010825in}}{\pgfqpoint{-0.014731in}{-0.014731in}}%
\pgfpathcurveto{\pgfqpoint{-0.010825in}{-0.018638in}}{\pgfqpoint{-0.005525in}{-0.020833in}}{\pgfqpoint{0.000000in}{-0.020833in}}%
\pgfpathlineto{\pgfqpoint{0.000000in}{-0.020833in}}%
\pgfpathclose%
\pgfusepath{stroke,fill}%
}%
\begin{pgfscope}%
\pgfsys@transformshift{0.723990in}{1.321093in}%
\pgfsys@useobject{currentmarker}{}%
\end{pgfscope}%
\begin{pgfscope}%
\pgfsys@transformshift{0.850607in}{1.208506in}%
\pgfsys@useobject{currentmarker}{}%
\end{pgfscope}%
\begin{pgfscope}%
\pgfsys@transformshift{0.977224in}{1.100038in}%
\pgfsys@useobject{currentmarker}{}%
\end{pgfscope}%
\begin{pgfscope}%
\pgfsys@transformshift{1.103841in}{1.035506in}%
\pgfsys@useobject{currentmarker}{}%
\end{pgfscope}%
\begin{pgfscope}%
\pgfsys@transformshift{1.230458in}{0.979213in}%
\pgfsys@useobject{currentmarker}{}%
\end{pgfscope}%
\begin{pgfscope}%
\pgfsys@transformshift{1.357075in}{0.932530in}%
\pgfsys@useobject{currentmarker}{}%
\end{pgfscope}%
\begin{pgfscope}%
\pgfsys@transformshift{1.483692in}{0.910562in}%
\pgfsys@useobject{currentmarker}{}%
\end{pgfscope}%
\begin{pgfscope}%
\pgfsys@transformshift{1.610310in}{0.887221in}%
\pgfsys@useobject{currentmarker}{}%
\end{pgfscope}%
\begin{pgfscope}%
\pgfsys@transformshift{1.736927in}{0.862507in}%
\pgfsys@useobject{currentmarker}{}%
\end{pgfscope}%
\begin{pgfscope}%
\pgfsys@transformshift{1.863544in}{0.840539in}%
\pgfsys@useobject{currentmarker}{}%
\end{pgfscope}%
\begin{pgfscope}%
\pgfsys@transformshift{1.990161in}{0.819943in}%
\pgfsys@useobject{currentmarker}{}%
\end{pgfscope}%
\begin{pgfscope}%
\pgfsys@transformshift{2.116778in}{0.806213in}%
\pgfsys@useobject{currentmarker}{}%
\end{pgfscope}%
\begin{pgfscope}%
\pgfsys@transformshift{2.243395in}{0.800721in}%
\pgfsys@useobject{currentmarker}{}%
\end{pgfscope}%
\begin{pgfscope}%
\pgfsys@transformshift{2.370012in}{0.791110in}%
\pgfsys@useobject{currentmarker}{}%
\end{pgfscope}%
\begin{pgfscope}%
\pgfsys@transformshift{2.496629in}{0.782872in}%
\pgfsys@useobject{currentmarker}{}%
\end{pgfscope}%
\begin{pgfscope}%
\pgfsys@transformshift{2.623246in}{0.774634in}%
\pgfsys@useobject{currentmarker}{}%
\end{pgfscope}%
\begin{pgfscope}%
\pgfsys@transformshift{2.749863in}{0.767769in}%
\pgfsys@useobject{currentmarker}{}%
\end{pgfscope}%
\begin{pgfscope}%
\pgfsys@transformshift{2.876480in}{0.763650in}%
\pgfsys@useobject{currentmarker}{}%
\end{pgfscope}%
\begin{pgfscope}%
\pgfsys@transformshift{3.003097in}{0.759531in}%
\pgfsys@useobject{currentmarker}{}%
\end{pgfscope}%
\begin{pgfscope}%
\pgfsys@transformshift{3.129714in}{0.755412in}%
\pgfsys@useobject{currentmarker}{}%
\end{pgfscope}%
\end{pgfscope}%
\begin{pgfscope}%
\pgfpathrectangle{\pgfqpoint{0.603704in}{0.549691in}}{\pgfqpoint{2.646296in}{1.300309in}}%
\pgfusepath{clip}%
\pgfsetrectcap%
\pgfsetroundjoin%
\pgfsetlinewidth{1.505625pt}%
\definecolor{currentstroke}{rgb}{1.000000,0.498039,0.054902}%
\pgfsetstrokecolor{currentstroke}%
\pgfsetdash{}{0pt}%
\pgfpathmoveto{\pgfqpoint{0.723990in}{0.608796in}}%
\pgfpathlineto{\pgfqpoint{0.850607in}{0.608944in}}%
\pgfpathlineto{\pgfqpoint{0.977224in}{0.609982in}}%
\pgfpathlineto{\pgfqpoint{1.103841in}{0.612058in}}%
\pgfpathlineto{\pgfqpoint{1.230458in}{0.615617in}}%
\pgfpathlineto{\pgfqpoint{1.357075in}{0.625107in}}%
\pgfpathlineto{\pgfqpoint{1.483692in}{0.634153in}}%
\pgfpathlineto{\pgfqpoint{1.610310in}{0.646461in}}%
\pgfpathlineto{\pgfqpoint{1.736927in}{0.664551in}}%
\pgfpathlineto{\pgfqpoint{1.863544in}{0.701326in}}%
\pgfpathlineto{\pgfqpoint{1.990161in}{0.723865in}}%
\pgfpathlineto{\pgfqpoint{2.116778in}{0.749964in}}%
\pgfpathlineto{\pgfqpoint{2.243395in}{0.774431in}}%
\pgfpathlineto{\pgfqpoint{2.370012in}{0.881789in}}%
\pgfpathlineto{\pgfqpoint{2.496629in}{0.874078in}}%
\pgfpathlineto{\pgfqpoint{2.623246in}{0.870371in}}%
\pgfpathlineto{\pgfqpoint{2.749863in}{0.961419in}}%
\pgfpathlineto{\pgfqpoint{2.876480in}{0.992114in}}%
\pgfpathlineto{\pgfqpoint{3.003097in}{1.004273in}}%
\pgfpathlineto{\pgfqpoint{3.129714in}{1.016432in}}%
\pgfusepath{stroke}%
\end{pgfscope}%
\begin{pgfscope}%
\pgfpathrectangle{\pgfqpoint{0.603704in}{0.549691in}}{\pgfqpoint{2.646296in}{1.300309in}}%
\pgfusepath{clip}%
\pgfsetbuttcap%
\pgfsetroundjoin%
\definecolor{currentfill}{rgb}{1.000000,0.498039,0.054902}%
\pgfsetfillcolor{currentfill}%
\pgfsetlinewidth{1.003750pt}%
\definecolor{currentstroke}{rgb}{1.000000,0.498039,0.054902}%
\pgfsetstrokecolor{currentstroke}%
\pgfsetdash{}{0pt}%
\pgfsys@defobject{currentmarker}{\pgfqpoint{-0.041667in}{-0.041667in}}{\pgfqpoint{0.041667in}{0.041667in}}{%
\pgfpathmoveto{\pgfqpoint{-0.041667in}{0.000000in}}%
\pgfpathlineto{\pgfqpoint{0.041667in}{0.000000in}}%
\pgfpathmoveto{\pgfqpoint{0.000000in}{-0.041667in}}%
\pgfpathlineto{\pgfqpoint{0.000000in}{0.041667in}}%
\pgfusepath{stroke,fill}%
}%
\begin{pgfscope}%
\pgfsys@transformshift{0.723990in}{0.608796in}%
\pgfsys@useobject{currentmarker}{}%
\end{pgfscope}%
\begin{pgfscope}%
\pgfsys@transformshift{0.850607in}{0.608944in}%
\pgfsys@useobject{currentmarker}{}%
\end{pgfscope}%
\begin{pgfscope}%
\pgfsys@transformshift{0.977224in}{0.609982in}%
\pgfsys@useobject{currentmarker}{}%
\end{pgfscope}%
\begin{pgfscope}%
\pgfsys@transformshift{1.103841in}{0.612058in}%
\pgfsys@useobject{currentmarker}{}%
\end{pgfscope}%
\begin{pgfscope}%
\pgfsys@transformshift{1.230458in}{0.615617in}%
\pgfsys@useobject{currentmarker}{}%
\end{pgfscope}%
\begin{pgfscope}%
\pgfsys@transformshift{1.357075in}{0.625107in}%
\pgfsys@useobject{currentmarker}{}%
\end{pgfscope}%
\begin{pgfscope}%
\pgfsys@transformshift{1.483692in}{0.634153in}%
\pgfsys@useobject{currentmarker}{}%
\end{pgfscope}%
\begin{pgfscope}%
\pgfsys@transformshift{1.610310in}{0.646461in}%
\pgfsys@useobject{currentmarker}{}%
\end{pgfscope}%
\begin{pgfscope}%
\pgfsys@transformshift{1.736927in}{0.664551in}%
\pgfsys@useobject{currentmarker}{}%
\end{pgfscope}%
\begin{pgfscope}%
\pgfsys@transformshift{1.863544in}{0.701326in}%
\pgfsys@useobject{currentmarker}{}%
\end{pgfscope}%
\begin{pgfscope}%
\pgfsys@transformshift{1.990161in}{0.723865in}%
\pgfsys@useobject{currentmarker}{}%
\end{pgfscope}%
\begin{pgfscope}%
\pgfsys@transformshift{2.116778in}{0.749964in}%
\pgfsys@useobject{currentmarker}{}%
\end{pgfscope}%
\begin{pgfscope}%
\pgfsys@transformshift{2.243395in}{0.774431in}%
\pgfsys@useobject{currentmarker}{}%
\end{pgfscope}%
\begin{pgfscope}%
\pgfsys@transformshift{2.370012in}{0.881789in}%
\pgfsys@useobject{currentmarker}{}%
\end{pgfscope}%
\begin{pgfscope}%
\pgfsys@transformshift{2.496629in}{0.874078in}%
\pgfsys@useobject{currentmarker}{}%
\end{pgfscope}%
\begin{pgfscope}%
\pgfsys@transformshift{2.623246in}{0.870371in}%
\pgfsys@useobject{currentmarker}{}%
\end{pgfscope}%
\begin{pgfscope}%
\pgfsys@transformshift{2.749863in}{0.961419in}%
\pgfsys@useobject{currentmarker}{}%
\end{pgfscope}%
\begin{pgfscope}%
\pgfsys@transformshift{2.876480in}{0.992114in}%
\pgfsys@useobject{currentmarker}{}%
\end{pgfscope}%
\begin{pgfscope}%
\pgfsys@transformshift{3.003097in}{1.004273in}%
\pgfsys@useobject{currentmarker}{}%
\end{pgfscope}%
\begin{pgfscope}%
\pgfsys@transformshift{3.129714in}{1.016432in}%
\pgfsys@useobject{currentmarker}{}%
\end{pgfscope}%
\end{pgfscope}%
\begin{pgfscope}%
\pgfpathrectangle{\pgfqpoint{0.603704in}{0.549691in}}{\pgfqpoint{2.646296in}{1.300309in}}%
\pgfusepath{clip}%
\pgfsetrectcap%
\pgfsetroundjoin%
\pgfsetlinewidth{1.505625pt}%
\definecolor{currentstroke}{rgb}{0.172549,0.627451,0.172549}%
\pgfsetstrokecolor{currentstroke}%
\pgfsetdash{}{0pt}%
\pgfpathmoveto{\pgfqpoint{0.723990in}{1.378605in}}%
\pgfpathlineto{\pgfqpoint{0.850607in}{1.491081in}}%
\pgfpathlineto{\pgfqpoint{0.977224in}{1.598823in}}%
\pgfpathlineto{\pgfqpoint{1.103841in}{1.661812in}}%
\pgfpathlineto{\pgfqpoint{1.230458in}{1.715301in}}%
\pgfpathlineto{\pgfqpoint{1.357075in}{1.754343in}}%
\pgfpathlineto{\pgfqpoint{1.483692in}{1.768862in}}%
\pgfpathlineto{\pgfqpoint{1.610310in}{1.781805in}}%
\pgfpathlineto{\pgfqpoint{1.736927in}{1.790895in}}%
\pgfpathlineto{\pgfqpoint{1.863544in}{1.781013in}}%
\pgfpathlineto{\pgfqpoint{1.990161in}{1.780993in}}%
\pgfpathlineto{\pgfqpoint{2.116778in}{1.771037in}}%
\pgfpathlineto{\pgfqpoint{2.243395in}{1.754709in}}%
\pgfpathlineto{\pgfqpoint{2.370012in}{1.669107in}}%
\pgfpathlineto{\pgfqpoint{2.496629in}{1.682631in}}%
\pgfpathlineto{\pgfqpoint{2.623246in}{1.692686in}}%
\pgfpathlineto{\pgfqpoint{2.749863in}{1.617070in}}%
\pgfpathlineto{\pgfqpoint{2.876480in}{1.592726in}}%
\pgfpathlineto{\pgfqpoint{3.003097in}{1.584858in}}%
\pgfpathlineto{\pgfqpoint{3.129714in}{1.576923in}}%
\pgfusepath{stroke}%
\end{pgfscope}%
\begin{pgfscope}%
\pgfpathrectangle{\pgfqpoint{0.603704in}{0.549691in}}{\pgfqpoint{2.646296in}{1.300309in}}%
\pgfusepath{clip}%
\pgfsetbuttcap%
\pgfsetroundjoin%
\definecolor{currentfill}{rgb}{0.172549,0.627451,0.172549}%
\pgfsetfillcolor{currentfill}%
\pgfsetlinewidth{1.003750pt}%
\definecolor{currentstroke}{rgb}{0.172549,0.627451,0.172549}%
\pgfsetstrokecolor{currentstroke}%
\pgfsetdash{}{0pt}%
\pgfsys@defobject{currentmarker}{\pgfqpoint{-0.033333in}{-0.041667in}}{\pgfqpoint{0.033333in}{0.020833in}}{%
\pgfpathmoveto{\pgfqpoint{0.000000in}{0.000000in}}%
\pgfpathlineto{\pgfqpoint{0.000000in}{-0.041667in}}%
\pgfpathmoveto{\pgfqpoint{0.000000in}{0.000000in}}%
\pgfpathlineto{\pgfqpoint{0.033333in}{0.020833in}}%
\pgfpathmoveto{\pgfqpoint{0.000000in}{0.000000in}}%
\pgfpathlineto{\pgfqpoint{-0.033333in}{0.020833in}}%
\pgfusepath{stroke,fill}%
}%
\begin{pgfscope}%
\pgfsys@transformshift{0.723990in}{1.378605in}%
\pgfsys@useobject{currentmarker}{}%
\end{pgfscope}%
\begin{pgfscope}%
\pgfsys@transformshift{0.850607in}{1.491081in}%
\pgfsys@useobject{currentmarker}{}%
\end{pgfscope}%
\begin{pgfscope}%
\pgfsys@transformshift{0.977224in}{1.598823in}%
\pgfsys@useobject{currentmarker}{}%
\end{pgfscope}%
\begin{pgfscope}%
\pgfsys@transformshift{1.103841in}{1.661812in}%
\pgfsys@useobject{currentmarker}{}%
\end{pgfscope}%
\begin{pgfscope}%
\pgfsys@transformshift{1.230458in}{1.715301in}%
\pgfsys@useobject{currentmarker}{}%
\end{pgfscope}%
\begin{pgfscope}%
\pgfsys@transformshift{1.357075in}{1.754343in}%
\pgfsys@useobject{currentmarker}{}%
\end{pgfscope}%
\begin{pgfscope}%
\pgfsys@transformshift{1.483692in}{1.768862in}%
\pgfsys@useobject{currentmarker}{}%
\end{pgfscope}%
\begin{pgfscope}%
\pgfsys@transformshift{1.610310in}{1.781805in}%
\pgfsys@useobject{currentmarker}{}%
\end{pgfscope}%
\begin{pgfscope}%
\pgfsys@transformshift{1.736927in}{1.790895in}%
\pgfsys@useobject{currentmarker}{}%
\end{pgfscope}%
\begin{pgfscope}%
\pgfsys@transformshift{1.863544in}{1.781013in}%
\pgfsys@useobject{currentmarker}{}%
\end{pgfscope}%
\begin{pgfscope}%
\pgfsys@transformshift{1.990161in}{1.780993in}%
\pgfsys@useobject{currentmarker}{}%
\end{pgfscope}%
\begin{pgfscope}%
\pgfsys@transformshift{2.116778in}{1.771037in}%
\pgfsys@useobject{currentmarker}{}%
\end{pgfscope}%
\begin{pgfscope}%
\pgfsys@transformshift{2.243395in}{1.754709in}%
\pgfsys@useobject{currentmarker}{}%
\end{pgfscope}%
\begin{pgfscope}%
\pgfsys@transformshift{2.370012in}{1.669107in}%
\pgfsys@useobject{currentmarker}{}%
\end{pgfscope}%
\begin{pgfscope}%
\pgfsys@transformshift{2.496629in}{1.682631in}%
\pgfsys@useobject{currentmarker}{}%
\end{pgfscope}%
\begin{pgfscope}%
\pgfsys@transformshift{2.623246in}{1.692686in}%
\pgfsys@useobject{currentmarker}{}%
\end{pgfscope}%
\begin{pgfscope}%
\pgfsys@transformshift{2.749863in}{1.617070in}%
\pgfsys@useobject{currentmarker}{}%
\end{pgfscope}%
\begin{pgfscope}%
\pgfsys@transformshift{2.876480in}{1.592726in}%
\pgfsys@useobject{currentmarker}{}%
\end{pgfscope}%
\begin{pgfscope}%
\pgfsys@transformshift{3.003097in}{1.584858in}%
\pgfsys@useobject{currentmarker}{}%
\end{pgfscope}%
\begin{pgfscope}%
\pgfsys@transformshift{3.129714in}{1.576923in}%
\pgfsys@useobject{currentmarker}{}%
\end{pgfscope}%
\end{pgfscope}%
\begin{pgfscope}%
\pgfsetrectcap%
\pgfsetmiterjoin%
\pgfsetlinewidth{0.803000pt}%
\definecolor{currentstroke}{rgb}{0.000000,0.000000,0.000000}%
\pgfsetstrokecolor{currentstroke}%
\pgfsetdash{}{0pt}%
\pgfpathmoveto{\pgfqpoint{0.603704in}{0.549691in}}%
\pgfpathlineto{\pgfqpoint{0.603704in}{1.850000in}}%
\pgfusepath{stroke}%
\end{pgfscope}%
\begin{pgfscope}%
\pgfsetrectcap%
\pgfsetmiterjoin%
\pgfsetlinewidth{0.803000pt}%
\definecolor{currentstroke}{rgb}{0.000000,0.000000,0.000000}%
\pgfsetstrokecolor{currentstroke}%
\pgfsetdash{}{0pt}%
\pgfpathmoveto{\pgfqpoint{3.250000in}{0.549691in}}%
\pgfpathlineto{\pgfqpoint{3.250000in}{1.850000in}}%
\pgfusepath{stroke}%
\end{pgfscope}%
\begin{pgfscope}%
\pgfsetrectcap%
\pgfsetmiterjoin%
\pgfsetlinewidth{0.803000pt}%
\definecolor{currentstroke}{rgb}{0.000000,0.000000,0.000000}%
\pgfsetstrokecolor{currentstroke}%
\pgfsetdash{}{0pt}%
\pgfpathmoveto{\pgfqpoint{0.603704in}{0.549691in}}%
\pgfpathlineto{\pgfqpoint{3.250000in}{0.549691in}}%
\pgfusepath{stroke}%
\end{pgfscope}%
\begin{pgfscope}%
\pgfsetrectcap%
\pgfsetmiterjoin%
\pgfsetlinewidth{0.803000pt}%
\definecolor{currentstroke}{rgb}{0.000000,0.000000,0.000000}%
\pgfsetstrokecolor{currentstroke}%
\pgfsetdash{}{0pt}%
\pgfpathmoveto{\pgfqpoint{0.603704in}{1.850000in}}%
\pgfpathlineto{\pgfqpoint{3.250000in}{1.850000in}}%
\pgfusepath{stroke}%
\end{pgfscope}%
\begin{pgfscope}%
\pgfsetbuttcap%
\pgfsetmiterjoin%
\definecolor{currentfill}{rgb}{1.000000,1.000000,1.000000}%
\pgfsetfillcolor{currentfill}%
\pgfsetfillopacity{0.800000}%
\pgfsetlinewidth{1.003750pt}%
\definecolor{currentstroke}{rgb}{0.800000,0.800000,0.800000}%
\pgfsetstrokecolor{currentstroke}%
\pgfsetstrokeopacity{0.800000}%
\pgfsetdash{}{0pt}%
\pgfpathmoveto{\pgfqpoint{0.910203in}{1.118080in}}%
\pgfpathlineto{\pgfqpoint{3.182528in}{1.118080in}}%
\pgfpathquadraticcurveto{\pgfqpoint{3.201806in}{1.118080in}}{\pgfqpoint{3.201806in}{1.137358in}}%
\pgfpathlineto{\pgfqpoint{3.201806in}{1.262333in}}%
\pgfpathquadraticcurveto{\pgfqpoint{3.201806in}{1.281611in}}{\pgfqpoint{3.182528in}{1.281611in}}%
\pgfpathlineto{\pgfqpoint{0.910203in}{1.281611in}}%
\pgfpathquadraticcurveto{\pgfqpoint{0.890926in}{1.281611in}}{\pgfqpoint{0.890926in}{1.262333in}}%
\pgfpathlineto{\pgfqpoint{0.890926in}{1.137358in}}%
\pgfpathquadraticcurveto{\pgfqpoint{0.890926in}{1.118080in}}{\pgfqpoint{0.910203in}{1.118080in}}%
\pgfpathlineto{\pgfqpoint{0.910203in}{1.118080in}}%
\pgfpathclose%
\pgfusepath{stroke,fill}%
\end{pgfscope}%
\begin{pgfscope}%
\pgfsetrectcap%
\pgfsetroundjoin%
\pgfsetlinewidth{1.505625pt}%
\definecolor{currentstroke}{rgb}{0.121569,0.466667,0.705882}%
\pgfsetstrokecolor{currentstroke}%
\pgfsetdash{}{0pt}%
\pgfpathmoveto{\pgfqpoint{0.929481in}{1.209276in}}%
\pgfpathlineto{\pgfqpoint{1.025870in}{1.209276in}}%
\pgfpathlineto{\pgfqpoint{1.122259in}{1.209276in}}%
\pgfusepath{stroke}%
\end{pgfscope}%
\begin{pgfscope}%
\pgfsetbuttcap%
\pgfsetroundjoin%
\definecolor{currentfill}{rgb}{0.121569,0.466667,0.705882}%
\pgfsetfillcolor{currentfill}%
\pgfsetlinewidth{1.003750pt}%
\definecolor{currentstroke}{rgb}{0.121569,0.466667,0.705882}%
\pgfsetstrokecolor{currentstroke}%
\pgfsetdash{}{0pt}%
\pgfsys@defobject{currentmarker}{\pgfqpoint{-0.020833in}{-0.020833in}}{\pgfqpoint{0.020833in}{0.020833in}}{%
\pgfpathmoveto{\pgfqpoint{0.000000in}{-0.020833in}}%
\pgfpathcurveto{\pgfqpoint{0.005525in}{-0.020833in}}{\pgfqpoint{0.010825in}{-0.018638in}}{\pgfqpoint{0.014731in}{-0.014731in}}%
\pgfpathcurveto{\pgfqpoint{0.018638in}{-0.010825in}}{\pgfqpoint{0.020833in}{-0.005525in}}{\pgfqpoint{0.020833in}{0.000000in}}%
\pgfpathcurveto{\pgfqpoint{0.020833in}{0.005525in}}{\pgfqpoint{0.018638in}{0.010825in}}{\pgfqpoint{0.014731in}{0.014731in}}%
\pgfpathcurveto{\pgfqpoint{0.010825in}{0.018638in}}{\pgfqpoint{0.005525in}{0.020833in}}{\pgfqpoint{0.000000in}{0.020833in}}%
\pgfpathcurveto{\pgfqpoint{-0.005525in}{0.020833in}}{\pgfqpoint{-0.010825in}{0.018638in}}{\pgfqpoint{-0.014731in}{0.014731in}}%
\pgfpathcurveto{\pgfqpoint{-0.018638in}{0.010825in}}{\pgfqpoint{-0.020833in}{0.005525in}}{\pgfqpoint{-0.020833in}{0.000000in}}%
\pgfpathcurveto{\pgfqpoint{-0.020833in}{-0.005525in}}{\pgfqpoint{-0.018638in}{-0.010825in}}{\pgfqpoint{-0.014731in}{-0.014731in}}%
\pgfpathcurveto{\pgfqpoint{-0.010825in}{-0.018638in}}{\pgfqpoint{-0.005525in}{-0.020833in}}{\pgfqpoint{0.000000in}{-0.020833in}}%
\pgfpathlineto{\pgfqpoint{0.000000in}{-0.020833in}}%
\pgfpathclose%
\pgfusepath{stroke,fill}%
}%
\begin{pgfscope}%
\pgfsys@transformshift{1.025870in}{1.209276in}%
\pgfsys@useobject{currentmarker}{}%
\end{pgfscope}%
\end{pgfscope}%
\begin{pgfscope}%
\definecolor{textcolor}{rgb}{0.000000,0.000000,0.000000}%
\pgfsetstrokecolor{textcolor}%
\pgfsetfillcolor{textcolor}%
\pgftext[x=1.199370in,y=1.175540in,left,base]{\color{textcolor}\rmfamily\fontsize{6.940000}{8.328000}\selectfont Sensors}%
\end{pgfscope}%
\begin{pgfscope}%
\pgfsetrectcap%
\pgfsetroundjoin%
\pgfsetlinewidth{1.505625pt}%
\definecolor{currentstroke}{rgb}{1.000000,0.498039,0.054902}%
\pgfsetstrokecolor{currentstroke}%
\pgfsetdash{}{0pt}%
\pgfpathmoveto{\pgfqpoint{1.750984in}{1.209276in}}%
\pgfpathlineto{\pgfqpoint{1.847373in}{1.209276in}}%
\pgfpathlineto{\pgfqpoint{1.943762in}{1.209276in}}%
\pgfusepath{stroke}%
\end{pgfscope}%
\begin{pgfscope}%
\pgfsetbuttcap%
\pgfsetroundjoin%
\definecolor{currentfill}{rgb}{1.000000,0.498039,0.054902}%
\pgfsetfillcolor{currentfill}%
\pgfsetlinewidth{1.003750pt}%
\definecolor{currentstroke}{rgb}{1.000000,0.498039,0.054902}%
\pgfsetstrokecolor{currentstroke}%
\pgfsetdash{}{0pt}%
\pgfsys@defobject{currentmarker}{\pgfqpoint{-0.041667in}{-0.041667in}}{\pgfqpoint{0.041667in}{0.041667in}}{%
\pgfpathmoveto{\pgfqpoint{-0.041667in}{0.000000in}}%
\pgfpathlineto{\pgfqpoint{0.041667in}{0.000000in}}%
\pgfpathmoveto{\pgfqpoint{0.000000in}{-0.041667in}}%
\pgfpathlineto{\pgfqpoint{0.000000in}{0.041667in}}%
\pgfusepath{stroke,fill}%
}%
\begin{pgfscope}%
\pgfsys@transformshift{1.847373in}{1.209276in}%
\pgfsys@useobject{currentmarker}{}%
\end{pgfscope}%
\end{pgfscope}%
\begin{pgfscope}%
\definecolor{textcolor}{rgb}{0.000000,0.000000,0.000000}%
\pgfsetstrokecolor{textcolor}%
\pgfsetfillcolor{textcolor}%
\pgftext[x=2.020873in,y=1.175540in,left,base]{\color{textcolor}\rmfamily\fontsize{6.940000}{8.328000}\selectfont Lost}%
\end{pgfscope}%
\begin{pgfscope}%
\pgfsetrectcap%
\pgfsetroundjoin%
\pgfsetlinewidth{1.505625pt}%
\definecolor{currentstroke}{rgb}{0.172549,0.627451,0.172549}%
\pgfsetstrokecolor{currentstroke}%
\pgfsetdash{}{0pt}%
\pgfpathmoveto{\pgfqpoint{2.425091in}{1.209276in}}%
\pgfpathlineto{\pgfqpoint{2.521480in}{1.209276in}}%
\pgfpathlineto{\pgfqpoint{2.617869in}{1.209276in}}%
\pgfusepath{stroke}%
\end{pgfscope}%
\begin{pgfscope}%
\pgfsetbuttcap%
\pgfsetroundjoin%
\definecolor{currentfill}{rgb}{0.172549,0.627451,0.172549}%
\pgfsetfillcolor{currentfill}%
\pgfsetlinewidth{1.003750pt}%
\definecolor{currentstroke}{rgb}{0.172549,0.627451,0.172549}%
\pgfsetstrokecolor{currentstroke}%
\pgfsetdash{}{0pt}%
\pgfsys@defobject{currentmarker}{\pgfqpoint{-0.033333in}{-0.041667in}}{\pgfqpoint{0.033333in}{0.020833in}}{%
\pgfpathmoveto{\pgfqpoint{0.000000in}{0.000000in}}%
\pgfpathlineto{\pgfqpoint{0.000000in}{-0.041667in}}%
\pgfpathmoveto{\pgfqpoint{0.000000in}{0.000000in}}%
\pgfpathlineto{\pgfqpoint{0.033333in}{0.020833in}}%
\pgfpathmoveto{\pgfqpoint{0.000000in}{0.000000in}}%
\pgfpathlineto{\pgfqpoint{-0.033333in}{0.020833in}}%
\pgfusepath{stroke,fill}%
}%
\begin{pgfscope}%
\pgfsys@transformshift{2.521480in}{1.209276in}%
\pgfsys@useobject{currentmarker}{}%
\end{pgfscope}%
\end{pgfscope}%
\begin{pgfscope}%
\definecolor{textcolor}{rgb}{0.000000,0.000000,0.000000}%
\pgfsetstrokecolor{textcolor}%
\pgfsetfillcolor{textcolor}%
\pgftext[x=2.694980in,y=1.175540in,left,base]{\color{textcolor}\rmfamily\fontsize{6.940000}{8.328000}\selectfont Efficiency}%
\end{pgfscope}%
\end{pgfpicture}%
\makeatother%
\endgroup%